\documentclass[sigconf,nonacm]{acmart}

\usepackage{booktabs}
\usepackage{tikz}
\usetikzlibrary{arrows.meta, positioning, calc}
\usepackage{pgfplots}
\pgfplotsset{compat=1.18}
\usepackage{xcolor}
\usepackage{listings}
\usepackage{balance}
\usepackage{algorithm}
\usepackage{algpseudocode}

\usepackage{enumitem}
\setlist{nosep,leftmargin=*}

\renewcommand\footnotetextcopyrightpermission[1]{}
\begin{document}

\title{TDAD: Test-Driven Agentic Development -- Reducing Code Regressions
  in AI Coding Agents via Graph-Based Impact Analysis}

\author{Pepe Alonso}
\affiliation{%
  \institution{Universidad ORT Uruguay}
  \country{}
}
\email{pepe@berkeley.edu}

\author{Sergio Yovine}
\affiliation{%
  \institution{Universidad ORT Uruguay}
  \country{}
}
\email{yovine@ort.edu.uy}

\author{Victor A. Braberman}
\affiliation{%
  \institution{DC, UBA, Argentina}
  \country{}
}
\email{vbraber@dc.uba.ar}

\begin{abstract}
AI coding agents can resolve real-world software issues, yet they frequently
introduce \emph{regressions}---breaking tests that previously passed.
Current benchmarks focus almost exclusively on resolution rate, leaving
regression behavior under-studied.

This work develops a method to enhance agents with appropriate structural
knowledge that leads to meaningful regression reduction while improving
resolution rate, and presents \textsc{TDAD} (Test-Driven Agentic
Development), an open-source tool that performs \emph{pre-change impact
analysis} for AI coding agents.  \textsc{TDAD} builds a dependency map between source code
and tests so that before committing a patch, the agent knows which tests
to verify and can self-correct.  The map is delivered as a lightweight
agent skill---a static text file the agent queries at runtime.
Evaluated on SWE-bench Verified with two open-weight models running on
consumer hardware (Qwen3-Coder~30B, 100~instances; Qwen3.5-35B-A3B,
25~instances), \textsc{TDAD} \textbf{reduced regressions by 70\%}
(6.08\%~$\to$~1.82\%) compared to a vanilla baseline.  In contrast,
adding TDD \emph{procedural} instructions without targeted test context
\emph{increased} regressions to 9.94\%---worse than no intervention at
all.  When deployed as an agent skill with a different model and framework,
\textsc{TDAD} \textbf{improved issue-resolution rate from 24\% to 32\%},
confirming that surfacing contextual information outperforms prescribing
procedural workflows.
All code, data, and logs are publicly available.
\end{abstract}

\maketitle

\section{Introduction}
\label{sec:intro}

Large-language-model (LLM) based coding agents have demonstrated impressive
ability to resolve real-world software issues.  On SWE-bench
Verified~\cite{jimenez2024swebench}, state-of-the-art agents resolve over 70\%
of GitHub issues from popular open-source repositories.  Evaluation, however,
centers almost exclusively on \emph{resolution rate}: the fraction of issues
for which the agent's patch passes the issue-specific tests.  A complementary
question---\emph{how many previously-passing tests does the patch
break?}---receives far less attention, despite recent evidence that
regressions and CI/CD failures are a leading cause of rejected agent-authored
pull requests~\cite{metr2026swebench,ehsani2026agentfail}.

Regressions are hard to prevent without dependency awareness.  An agent can
either run \emph{all} tests (too slow for large codebases) or run only tests
near the changed files (missing indirect dependencies).  This problem
parallels classical \emph{regression test selection}
(RTS)~\cite{elbaum2014techniques,legunsen2016extensive}, but the agentic
context introduces novel requirements: changes are generated
programmatically, the agent has a limited context window, and test
verification must happen \emph{before} submission rather than in a CI
pipeline afterward.  In the baseline experiments, a vanilla agent caused
562~pass-to-pass (P2P) test failures across 100~instances---an average
of 6.5~broken tests per generated patch.

The SWE-bench evaluation harness already collects the data needed to measure
regressions: the \texttt{PASS\_TO\_PASS} test set records tests that passed
before the gold patch and should still pass afterward.  Yet this data is not
surfaced in leaderboard rankings.  Recent evidence reinforces this concern:
METR~\cite{metr2026swebench} found that roughly half of SWE-bench-passing
patches would not be merged by real maintainers, and Ehsani et
al.~\cite{ehsani2026agentfail} show that CI/CD failures are a leading cause
of rejected agent-authored pull requests.  Regression rate deserves
first-class status alongside resolution rate.

\begin{figure}[b]
\centering
\resizebox{\columnwidth}{!}{%
\begin{tikzpicture}[
    node distance=0.25cm,
    wfbox/.style={draw, rounded corners, minimum height=0.65cm,
                  minimum width=1.2cm, align=center, font=\scriptsize},
    wftdad/.style={wfbox, fill=orange!20, draw=orange!60!black,
                   font=\scriptsize\bfseries},
    wfarr/.style={-{Stealth[length=3pt]}, semithick},
    wflarr/.style={-{Stealth[length=3pt]}, semithick, dashed, red!60!black},
  ]
  \node[wfbox, fill=blue!8] (issue) {Read\\[-1pt]issue};
  \node[wfbox, fill=blue!8, right=0.25cm of issue] (plan) {Plan\\[-1pt]fix};
  \node[wfbox, fill=blue!8, right=0.25cm of plan] (impl) {Implement};
  \node[wftdad, right=0.25cm of impl] (tdadnode) {TDAD:\\[-1pt]at-risk tests};
  \node[wfbox, fill=green!12, right=0.25cm of tdadnode] (verify) {Verify\\[-1pt]tests};
  \node[wfbox, fill=green!12, right=0.25cm of verify] (patch) {Submit\\[-1pt]patch};

  \draw[wfarr] (issue) -- (plan);
  \draw[wfarr] (plan) -- (impl);
  \draw[wfarr] (impl) -- (tdadnode);
  \draw[wfarr] (tdadnode) -- (verify);
  \draw[wfarr] (verify) -- (patch);

  \draw[wflarr] (verify.south) -- ++(0,-0.4) -| node[below, pos=0.25,
    font=\tiny, red!60!black] {fix regressions} (impl.south);
\end{tikzpicture}%
}
\caption{Agentic workflow with \textsc{TDAD}.  After implementing changes,
  the agent queries \textsc{TDAD}'s test map to identify at-risk tests,
  verifies them, and self-corrects if regressions are detected.  The dashed
  arrow represents the self-correction loop.}
\label{fig:workflow}
\end{figure}
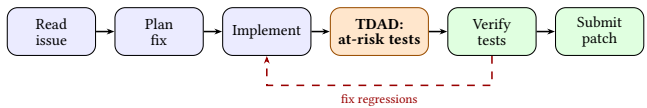

This paper presents \textsc{TDAD} (\textbf{T}est-\textbf{D}riven
\textbf{A}gentic \textbf{D}evelopment), an open-source tool that performs
\emph{pre-change impact analysis} for AI coding agents.  \textsc{TDAD}
builds a dependency map between source code and tests so that before
committing a patch, the agent knows exactly which tests to verify and can
self-correct (Figure~\ref{fig:workflow}).  The key insight is that agents
do not need to be told \emph{how} to do TDD; they need to be told
\emph{which tests to check}.  At runtime the agent needs only \texttt{grep}
and \texttt{pytest}---no graph database, MCP server, or API calls.

The evaluation spans two phases on SWE-bench Verified---a controlled
comparison on 100~instances with a single model, followed by a
generalization study with a different model and agent framework---yielding
four key findings:

\begin{enumerate}
  \item \textbf{70\% regression reduction.}
    In Phase~1 (100~instances, Qwen3-Coder~30B, single-agent setup),
    \textsc{TDAD} reduced test-level regression rate from
    6.08\% to 1.82\% (562~$\to$~155 P2P failures).
  \item \textbf{TDD Prompting Paradox.}
    In the same Phase~1 setup, an ablation adding TDD \emph{procedural}
    instructions (write tests first, then implement) \emph{without}
    telling the agent \emph{which specific tests} to check actually
    \emph{increased} regressions to 9.94\%---worse than vanilla.
    Procedural instructions without targeted context are counterproductive.
  \item \textbf{Resolution improvement as agent skill.}
    In Phase~2 (25~instances, different model Qwen3.5-35B-A3B + OpenCode agent),
    \textsc{TDAD} improved resolution from 24\% to 32\% and generation
    from 40\% to 68\%, confirming generalization beyond the original setup.
  \item \textbf{Autonomous self-improvement.}
    An auto-improvement loop (10-instance subset) iteratively refined
    \textsc{TDAD}'s own configuration, raising resolution from 12\% to
    60\% with 0\% regression.
\end{enumerate}

\noindent\textbf{Contributions.}\enspace
(1)~An open-source tool for graph-based test impact analysis
(\texttt{pip install tdad});
(2)~a benchmark methodology elevating regression rate to a first-class metric;
(3)~empirical evidence across two models and agent frameworks;
(4)~an autonomous auto-improvement loop for iterative tool refinement;
(5)~all code, data, and 28~experiments released under MIT license.

\section{Related Work}
\label{sec:related}

\subsection{AI Coding Agents and Benchmarks}

SWE-bench~\cite{jimenez2024swebench} evaluates agents on resolving GitHub
issues from 12~popular Python repositories, establishing the primary benchmark
for AI coding agents.  The SWE-bench Verified subset provides human-validated
ground truth for 500~instances.  SWE-Agent~\cite{yang2024sweagent} introduces
an agent--computer interface optimized for repository navigation, achieving
strong results through careful tool design.
AutoCodeRover~\cite{zhang2024autocoderover} combines code search with
spectrum-based fault localization to improve issue resolution.
OpenHands~\cite{wang2024openhands} provides a unified platform for developing
and evaluating coding agents across multiple benchmarks.

The benchmark ecosystem continues to evolve.
SWE-smith~\cite{yang2025swesmith} scales training data by automatically
synthesizing task instances from arbitrary Python codebases, producing
50k~instances and training open-source models that achieve 40.2\% on
SWE-bench Verified.
SWE-Bench++~\cite{wang2025swebenchpp} extends evaluation to 11~programming
languages with 11k~instances.
SWE-CI~\cite{chen2026sweci} shifts focus from one-shot bug fixing to
long-term codebase maintenance via continuous integration loops, requiring
agents to handle sequences of commits over months.

All of these efforts focus primarily on resolution rate; regression behavior
is reported only incidentally, if at all.  The SWE-bench harness \emph{does}
execute P2P tests, but results are not surfaced in leaderboard rankings.
Ehsani et al.~\cite{ehsani2026agentfail} study 33k agent-authored pull
requests on GitHub and find that CI/CD pipeline failures and regression
are among the most common rejection reasons, confirming that resolution
alone is an insufficient measure.
This work argues that this omission creates a perverse incentive that rewards
aggressive patching regardless of side effects.

\subsection{Regression Testing}

Regression test selection (RTS) and prioritization have a long history in
software engineering.  Elbaum et al.~\cite{elbaum2014techniques} survey
techniques for improving regression testing in CI environments, establishing
that even simple selection strategies can dramatically reduce test execution
time.  Legunsen et al.~\cite{legunsen2016extensive} evaluate static RTS at
scale, finding that class-level dependency tracking achieves good precision.
Gligoric et al.~\cite{gligoric2015practical} propose dynamic file-level
dependency tracking via filesystem monitoring.
Chianti~\cite{ren2004chianti} performs method-level change impact analysis
using call-graph differencing for Java programs.

\textsc{TDAD} adapts these classical RTS ideas to a novel setting: AI coding
agents.  The differences are threefold.
\emph{Timing}: traditional RTS optimizes CI time by selecting which tests to
run \emph{after} a change is committed; \textsc{TDAD} operates \emph{before}
the agent commits, enabling self-correction rather than post-hoc detection.
\emph{Consumer}: the output of traditional RTS feeds a test runner or CI
pipeline; \textsc{TDAD}'s output is a static text file designed to be consumed
by an LLM agent via \texttt{grep}, respecting context-window limits.
\emph{Goal}: traditional RTS minimizes execution time for a fixed test suite;
\textsc{TDAD} minimizes \emph{regressions} by surfacing which tests the agent
should verify, not merely which tests to run.

\subsection{Graph-Based Code Analysis}

Code property graphs~\cite{yamaguchi2014modeling} unify ASTs,
control-flow graphs, and program-dependence graphs for vulnerability
detection.  GraphRAG~\cite{edge2024graphrag} demonstrates that
graph-structured retrieval outperforms flat vector search for complex
reasoning.  GRACE~\cite{wang2025grace} constructs a multi-level code
graph (file structure, AST, call graph, class hierarchy) for
repository-aware code completion, achieving 8\% improvement over
prior graph-based RAG baselines.  \textsc{TDAD} applies a similar
graph-first principle but targets a different task: rather than
code completion, it traverses an explicit code--test dependency graph
to identify impacted tests with high precision.

\subsection{TDD and AI Agents}

Test-driven development (TDD)~\cite{beck2003tdd} prescribes writing tests
before implementation code, creating a tight feedback loop that catches
regressions early.  \textsc{TDAD} borrows TDD's core \emph{philosophy}---verify
before you commit---but replaces TDD's \emph{procedure} (write test $\to$ red
$\to$ green $\to$ refactor) with \emph{contextual guidance}: rather than
instructing the agent to follow TDD steps, \textsc{TDAD} tells the agent
\emph{which specific tests} are at risk.  This distinction directly motivates
the TDD Prompting Paradox (\S\ref{sec:paradox}), where procedural TDD
instructions without contextual information actually increased regressions.
Cui~\cite{cui2025testsasprompt} introduces a TDD benchmark where test cases
serve as both prompt and verification, finding that \emph{instruction following}
and in-context learning matter more than general coding proficiency for TDD
success, and that performance degrades when instructions are long---a result
consistent with the present work's finding that smaller models benefit more
from concise, targeted context than from lengthy procedural instructions.

Rehan~\cite{rehan2026tdad} independently proposes a ``Test-Driven AI Agent
Definition'' framework that compiles agent \emph{prompts} from behavioral
specifications via iterative test--refine loops, achieving 97.2\% regression
safety.  While sharing the TDD philosophy and even the TDAD acronym, the two
works target different problems: Rehan validates TDD for agent \emph{behavior
compliance}; this paper validates it for agent-generated \emph{code patches}.

\section{System Design}
\label{sec:design}

\subsection{Architecture Overview}

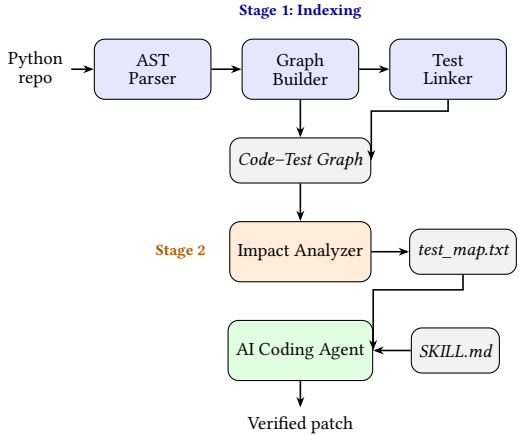
\begin{figure}[t]
\centering
\begin{tikzpicture}[
    node distance=0.35cm and 0.5cm,
    box/.style={draw, rounded corners, minimum height=0.8cm,
                minimum width=1.55cm, align=center, font=\footnotesize},
    data/.style={draw, rounded corners, fill=gray!10,
                 minimum height=0.6cm, align=center,
                 font=\footnotesize\itshape},
    arr/.style={-{Stealth[length=4pt]}, semithick},
  ]
  \node[box, fill=blue!10] (parser) {AST\\[-1pt]Parser};
  \node[box, fill=blue!10, right=0.4cm of parser] (builder) {Graph\\[-1pt]Builder};
  \node[box, fill=blue!10, right=0.4cm of builder] (linker) {Test\\[-1pt]Linker};

  \node[data, below=0.5cm of builder] (graph) {Code--Test Graph};

  \node[box, fill=orange!15, below=0.5cm of graph] (impact) {Impact Analyzer};
  \node[data, right=0.5cm of impact] (testmap) {test\_map.txt};

  \node[box, fill=green!12, below=0.5cm of impact] (agent) {AI Coding Agent};
  \node[data, right=0.5cm of agent] (skill) {SKILL.md};

  \draw[arr] (parser) -- (builder);
  \draw[arr] (builder) -- (linker);
  \draw[arr] (builder.south) -- (graph);
  \draw[arr] (linker.south) -- ++(0,-0.15) -| (graph.east);
  \draw[arr] (graph) -- (impact);
  \draw[arr] (impact) -- (testmap);
  \draw[arr] (testmap.south) -- ++(0,-0.2) -| (agent.east);
  \draw[arr] (skill) -- (agent);

  \node[left=0.3cm of parser, font=\footnotesize, align=center]
       (repo) {Python\\[-1pt]repo};
  \draw[arr] (repo) -- (parser);
  \node[below=0.35cm of agent, font=\footnotesize] (patch) {Verified patch};
  \draw[arr] (agent) -- (patch);

  \node[above=0.15cm of builder, font=\scriptsize\bfseries, blue!60!black]
       {Stage 1: Indexing};
  \node[left=0.1cm of impact, font=\scriptsize\bfseries, orange!70!black,
        anchor=east, xshift=-0.1cm] {Stage 2};
\end{tikzpicture}
\caption{\textsc{TDAD} pipeline.  Stage~1 indexes the repository into a
  code--test graph.  Stage~2 exports impacted tests as a static test map.
  The agent uses \texttt{grep} on the test map to find tests to verify.}
\label{fig:architecture}
\end{figure}

Figure~\ref{fig:architecture} shows the \textsc{TDAD} pipeline.  Stage~1
(\emph{indexing}) parses a Python repository and builds a code--test
dependency graph.  Stage~2 (\emph{impact analysis}) identifies tests affected
by changed files and exports a static text file (\texttt{test\_map.txt}).
The agent receives this file plus a 20-line skill definition; no MCP server,
API calls, or graph database is required at runtime.

\subsection{Graph Schema}

The graph uses four node types and five edge types
(Table~\ref{tab:schema}).  Nodes represent structural units; edges capture
static relationships from AST analysis.

\begin{table}[t]
\caption{Graph schema: node and edge types.}
\label{tab:schema}
\small
\begin{tabular}{@{}lll@{}}
\toprule
\textbf{Type} & \textbf{Entity} & \textbf{Key attributes} \\
\midrule
\multicolumn{3}{@{}l}{\emph{Node types}} \\
File     & Python source file    & path, content\_hash \\
Function & Top-level function    & name, file, lines, signature \\
Class    & Class definition      & name, file, bases \\
Test     & Test function/method  & name, file, is\_test \\
\midrule
\multicolumn{3}{@{}l}{\emph{Edge types}} \\
CONTAINS  & File $\to$ Function/Class  & structural containment \\
CALLS     & Function $\to$ Function    & static call resolution \\
IMPORTS   & File $\to$ File            & import tracking \\
TESTS     & Test $\to$ Function/Class  & test--code linkage \\
INHERITS  & Class $\to$ Class          & base-class relationship \\
\bottomrule
\end{tabular}
\end{table}

\subsection{Indexing Pipeline}

The indexer has three components, each producing a specific subset of graph
elements.

\paragraph{AST Parser.}
Parses each Python file using the standard-library \texttt{ast} module.
Extracts: (a)~function definitions with signatures, line ranges, and
docstrings; (b)~class definitions with base classes and methods; (c)~import
statements; and (d)~call targets within function bodies, resolved via a
recursive visitor that handles both simple names
and attribute chains.  Test files are identified by naming conventions:
\texttt{test\_*.py} and \texttt{*\_test.py} for files,
\texttt{test\_*} for functions, and \texttt{Test*} for classes.

\paragraph{Graph Builder.}
Populates nodes and edges.  Each file produces a \textsc{File} node and child
\textsc{Function}/\textsc{Class} nodes connected via \textsc{CONTAINS} edges.
Call targets are resolved using module-scoped name resolution, creating
\textsc{CALLS} edges.  Import statements produce \textsc{IMPORTS} edges;
class inheritance produces \textsc{INHERITS} edges.

\paragraph{Test Linker.}
Creates \textsc{TESTS} edges connecting test nodes to the code they
exercise.  This is the most critical component, as Python projects use
diverse conventions for organizing tests.  Three strategies in priority
order:
(1)~\emph{naming conventions}
(\texttt{test\_foo.py}~$\to$~\texttt{foo.py}, including
underscore-prefixed variants common in scientific Python);
(2)~\emph{prefix matching} (progressive truncation of test-file stems
to find the best source-file match);
(3)~\emph{directory proximity} for disambiguation when a stem matches
multiple source files.  For monolithic test modules
(e.g., Django's \texttt{tests.py}), a dedicated proximity algorithm
maps to source files in the nearest non-test ancestor.

\subsection{Impact Analysis}

Given changed files, four strategies run in parallel
(Table~\ref{tab:strategies}) and scores are merged:
\begin{equation}
  \text{score} = (1 - c_w) \cdot w_{\text{strategy}} + c_w \cdot
  \text{confidence}
  \label{eq:score}
\end{equation}
where $c_w = 0.3$ is the confidence weight and $\text{confidence} \in [0,1]$
reflects link strength (1.0 for direct \textsc{TESTS} edges, 0.56 for
transitive call chains, 0.5 for coverage, 0.45 for imports).  When a test
appears via multiple strategies, only the highest score is kept.  Tests are
selected in three tiers: high ($\geq 0.8$), medium ($0.5$--$0.8$), and low
($< 0.5$), up to a configurable maximum (default: 50).  Three weight profiles
are available: \emph{conservative} (favors precision), \emph{balanced}
(default), and \emph{aggressive} (favors recall).

\smallskip\noindent\textbf{Justification of constants.}\enspace
The confidence weight $c_w = 0.3$ allocates 70\% of the score to
strategy-specific weights and 30\% to link-quality confidence, ensuring
that the \emph{type} of relationship (direct test, transitive call chain,
coverage, import) dominates while still rewarding stronger evidence.
The confidence values themselves---1.0, 0.56, 0.5, 0.45---form a
monotonically decreasing scale reflecting trust in link quality: direct
\textsc{TESTS} edges are ground-truth associations, transitive call chains
decay with hop distance, file-level coverage is coarser, and import-level
links are the weakest signal.  These values were set heuristically from
the graph structure and subsequently refined through the autonomous
auto-improvement loop (\S\ref{sec:autoloop}), where the agent itself
iterated on weight configurations to minimize regressions.  The three
weight profiles (conservative, balanced, aggressive) allow practitioners
to adapt to different codebases without modifying the formula.

\begin{table}[t]
\caption{Impact analysis strategies and base weights (balanced profile).}
\label{tab:strategies}
\small
\begin{tabular}{@{}llp{4cm}@{}}
\toprule
\textbf{Strategy} & \textbf{Weight} & \textbf{Description} \\
\midrule
Direct     & 0.95 & Directly tests changed code \\
Transitive & 0.70 & 1--3 call-chain hops to changed code \\
Coverage   & 0.80 & File-level dependency \\
Imports    & 0.50 & Imports changed files \\
\bottomrule
\end{tabular}
\end{table}

\subsection{Agent Integration: TDAD as a Skill}

\textsc{TDAD} integrates with agents through two static artifacts.
\textbf{test\_map.txt} maps source files to test files (one line per
mapping, \texttt{grep}-able).  \textbf{SKILL.md} is a 20-line
definition: (1)~fix the bug, (2)~\texttt{grep}
\texttt{test\_map.txt} for related tests, (3)~run them and fix
failures.  This brevity proved critical: through the auto-improvement
loop (\S\ref{sec:autoloop}), we found that simplifying SKILL.md from
107 to 20~lines alone quadrupled resolution from 12\% to 50\%.

\subsection{Backend Architecture}

\textsc{TDAD} originally used Neo4j with Docker for graph storage.
Through iterative development, we migrated to a \textbf{NetworkX}
in-memory backend as the default, eliminating the Docker dependency
entirely.  Installation is now \texttt{pip install tdad} with zero
external requirements.  Graph persistence uses pickle
(\texttt{.tdad/graph.pkl}).  Neo4j remains available via
\texttt{TDAD\_BACKEND=neo4j} for large-scale deployments.  Both
backends expose the same \texttt{GraphDB} interface.

\section{Experimental Setup}
\label{sec:setup}

\subsection{Benchmark}

The evaluation uses SWE-bench Verified~\cite{jimenez2024swebench}, a human-validated
subset of SWE-bench containing 500~GitHub issues from 12~popular Python
repositories (Django, scikit-learn, sympy, matplotlib, astropy, Flask, pytest,
and others).  Each instance specifies: (a)~a GitHub issue description,
(b)~a repository snapshot at issue creation time, (c)~\texttt{FAIL\_TO\_PASS}
tests that the gold patch fixes, and (d)~\texttt{PASS\_TO\_PASS} tests that
should remain passing.  Phase~1 uses 100~instances (first 100 in canonical
order); Phase~2 uses 25~instances selected for diversity.

\subsection{Models and Agents}

Two experimental phases were conducted (Table~\ref{tab:configs}):

\textbf{Phase~1} (100~instances) uses \textbf{Qwen3-Coder~30B}~\cite{qwen2025qwen3coder} (Q4\_K\_M 4-bit quantization) served via llama.cpp on
consumer hardware, with a 32K-token context window and temperature~0 for
deterministic outputs.  Each instance was given up to 15~minutes for patch
generation.  This phase evaluates regression reduction across three prompt
configurations.  We chose a smaller, locally-run model for three reasons:
(1)~to demonstrate that \textsc{TDAD}'s benefits are not contingent on
frontier-scale models; (2)~to enable fully reproducible experiments without
API costs; and (3)~to stress-test the approach where every context token
matters.

\textbf{Phase~2} (25~instances) uses \textbf{Qwen3.5-35B-A3B} (4-bit
quantization, mixture-of-experts with 3B active parameters) served via MLX
on Apple Silicon, using the \textbf{OpenCode~v1.2.24} coding agent.  This
phase evaluates \textsc{TDAD} deployed as a reusable agent skill with the
NetworkX backend, using a different model, quantization framework, and agent
from Phase~1 to test generalization of the approach.

\paragraph{Baseline context.}
For reference, Qwen3-Coder~30B achieves ${\sim}$52\% on SWE-bench Verified
with full-featured scaffolds (SWE-Agent, 100+~turns), and Qwen3.5-35B-A3B
achieves 69.2\% as reported by Qwen.  The lower baselines in this work
(31\% and 24\%) reflect 4-bit quantization, consumer hardware, shorter
context windows, and simpler agent scaffolds---conditions chosen to
stress-test \textsc{TDAD} under resource constraints.

\begin{table}[t]
\caption{Experimental configurations across both phases.}
\label{tab:configs}
\small
\begin{tabular}{@{}llp{3.8cm}@{}}
\toprule
\textbf{Phase} & \textbf{Config} & \textbf{Description} \\
\midrule
1 & Vanilla     & Default prompt, no TDD or graph \\
1 & TDD Prompt  & Adds TDD workflow instructions \\
1 & GraphRAG+TDD & TDD + SKILL.md + test\_map.txt \\
\midrule
2 & Baseline    & OpenCode agent, no TDAD skill \\
2 & TDAD Skill  & OpenCode + TDAD skill + NetworkX \\
\bottomrule
\end{tabular}
\end{table}

\subsection{Evaluation Protocol}

Patches are evaluated using the SWE-bench Docker harness, which for each
instance: (1)~checks out the repository at the issue's base commit;
(2)~applies the agent's patch; (3)~runs fail-to-pass (F2P) tests (resolution);
(4)~runs P2P tests (regression detection).  Each instance is evaluated in
an isolated Docker container.  Empty patches are excluded from P2P evaluation.

Four metrics are reported:
\begin{itemize}
  \item \textbf{Resolution rate:} fraction of instances where all F2P tests
    pass.
  \item \textbf{Generation rate:} fraction producing a non-empty patch.
  \item \textbf{Test-level regression rate:} total P2P failures / total P2P
    tests across all instances.  This is the primary regression metric because
    it distinguishes a patch breaking 1~test from one breaking 100.
  \item \textbf{Instance-level regression rate:} fraction of generated patches
    with $\geq 1$ P2P failure.
\end{itemize}

\section{Results and Analysis}
\label{sec:results}

\subsection{Phase 1: Regression Reduction (100 Instances)}

Table~\ref{tab:main} presents the Phase~1 results with Qwen3-Coder~30B.

\begin{table}[t]
\caption{Phase~1 results: SWE-bench Verified, 100 instances, Qwen3-Coder~30B.
  $\downarrow$ = lower is better.}
\label{tab:main}
\small
\begin{tabular}{@{}lrrr@{}}
\toprule
\textbf{Metric} & \textbf{Vanilla} & \textbf{TDD} & \textbf{GraphRAG} \\
 & & \textbf{Prompt} & \textbf{+ TDD} \\
\midrule
Resolution Rate            & \textbf{31\%}  & \textbf{31\%}  & 29\% \\
Generation Rate            & \textbf{86\%}  & 75\%           & 74\% \\
\midrule
Total P2P Tests            & 9{,}245        & 8{,}040        & 8{,}536 \\
P2P Failures $\downarrow$  & 562            & 799            & \textbf{155} \\
Test Regr.\ $\downarrow$   & 6.08\%         & 9.94\%         & \textbf{1.82\%} \\
Inst.\ Regr.\ $\downarrow$ & \textbf{30.2\%} & 33.3\%       & 33.3\% \\
Catastrophic$^*$ $\downarrow$ & 3           & 5              & \textbf{1} \\
\bottomrule
\multicolumn{4}{@{}l}{\footnotesize $^*$Instances where all P2P tests failed.}
\end{tabular}
\end{table}

GraphRAG+TDD achieved a \textbf{72\% reduction} in P2P failures
(562~$\to$~155) and a \textbf{70\% reduction in regression rate}
(6.08\%~$\to$~1.82\%).  Compared to TDD-only, the reduction is even
larger: 81\% fewer failures (799~$\to$~155).  Catastrophic regressions
(all P2P tests failing) dropped from 3 in vanilla and 5 in TDD-only
to just~1 in GraphRAG+TDD.

Resolution decreased modestly ($-2$~pp), driven by a higher empty-patch rate
(26\% vs.\ 14\%); the agent abstained more often when the test map indicated
risk.  Among patches that were actually generated, GraphRAG was no less likely
to resolve the issue.

\paragraph{Note on denominators and instance-level rates.}
The three configurations produced different numbers of non-empty patches
(86, 75, and 74 respectively), yielding different P2P test pools
(9{,}245 vs.\ 8{,}040 vs.\ 8{,}536 total tests).  The instance-level
regression rates (30.2\%, 33.3\%, 33.3\%) appear to contradict the
test-level result; however, in absolute terms GraphRAG produced
\emph{fewer} regressing instances (25 of 74) than vanilla (26 of 86).
The higher \emph{rate} is a denominator artifact of GraphRAG's lower
generation rate.  We report both metrics for transparency but emphasize
test-level regression rate as primary because it captures regression
\emph{severity}: a patch breaking 322~tests is qualitatively different
from one breaking~1.

Table~\ref{tab:catastrophic} illustrates regression severity on
specific instances.  On \texttt{astropy\hbox{-}13977} (322~P2P tests),
vanilla failed all~322 while GraphRAG failed only~12.  On
\texttt{django\hbox{-}13089}, TDD prompting caused total failure
(352/352) while vanilla failed only~4.  These cases show that TDD
prompting without localization can be \emph{catastrophically} worse
than doing nothing.

\begin{table}[t]
\caption{Selected instances with severe regressions.  Numbers = P2P failures /
  total.  ``---'' = empty patch (not evaluated).}
\label{tab:catastrophic}
\small
\begin{tabular}{@{}lrrr@{}}
\toprule
\textbf{Instance} & \textbf{Vanilla} & \textbf{TDD} & \textbf{GraphRAG} \\
\midrule
astropy-13977 (322)   & 322/322 & 4/322   & 12/322 \\
astropy-8872 (80)     & 80/80   & 80/80   & --- \\
django-13089 (352)    & 4/352   & 352/352 & --- \\
django-11532 (148)    & ---     & 148/148 & --- \\
django-11299 (103)    & ---     & 103/103 & --- \\
\bottomrule
\end{tabular}
\end{table}

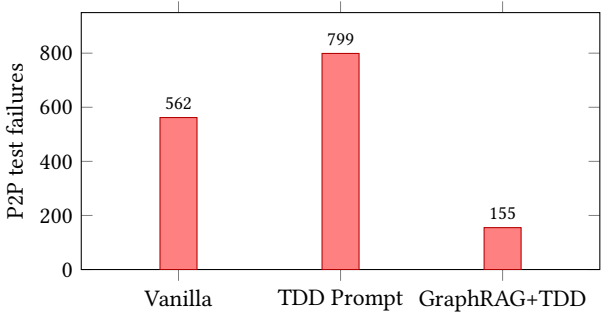
\begin{figure}[t]
\centering
\begin{tikzpicture}
\begin{axis}[
    ybar,
    width=\columnwidth,
    height=5cm,
    bar width=14pt,
    ylabel={P2P test failures},
    symbolic x coords={Vanilla, TDD Prompt, GraphRAG+TDD},
    xtick=data,
    ymin=0, ymax=950,
    nodes near coords,
    nodes near coords align={vertical},
    every node near coord/.append style={font=\footnotesize\bfseries},
    enlarge x limits=0.3,
]
\addplot[fill=red!50!white, draw=red!70!black] coordinates {
    (Vanilla, 562) (TDD Prompt, 799) (GraphRAG+TDD, 155)
};
\end{axis}
\end{tikzpicture}
\caption{Total P2P failures per approach (Phase~1).  GraphRAG+TDD reduces
  failure count by 72\% vs.\ vanilla and 81\% vs.\ TDD-only.}
\label{fig:p2p}
\end{figure}

\subsection{The TDD Prompting Paradox}
\label{sec:paradox}

TDD prompting \emph{without} graph context increased regressions
(6.08\%~$\to$~9.94\%), producing 42\% more P2P failures than
vanilla.  Two factors explain this:

\paragraph{Verbose prompts hurt small models.}
The TDD prompt consumed context tokens with procedural instructions, pushing
out repository context that the 30B model needed for accurate changes.

\paragraph{Ambition without localization.}
TDD-prompted agents attempted more ambitious fixes, touching more files.
Without graph-based knowledge of \emph{which} tests to verify, this ambition
caused collateral damage.  The five catastrophic regressions in TDD-only
(vs.\ three in vanilla) confirm this.  GraphRAG counteracted both effects:
the 20-line SKILL.md freed context tokens, and the test map focused attention
on tests at risk.

\paragraph{Ablation evidence.}
Prior experiments isolated the respective contributions.  Shortening the
prompt \emph{without} graph context (from 119 to 49~lines) decreased
resolution from 30\% to 20\%, confirming that prompt brevity alone does
not explain GraphRAG's improvement.  Conversely, doubling prompt length
without graph context (49-line vanilla vs.\ 119-line TDD) produced
identical 31\% resolution, showing that additional procedural text neither
helped nor hurt in isolation.  The improvement requires graph-derived
context: a short prompt \emph{plus} the test map.

\textbf{Practical implication:} for smaller models, context (which tests to
check) outperforms procedure (how to do TDD).

\subsection{Resolution vs.\ Regression Trade-off}

The modest 2-percentage-point resolution drop (31\%~$\to$~29\%) represents an
intentional conservative bias.  GraphRAG agents produced more empty patches
(26\% vs.\ 14\%) when the test map indicated high regression risk; the agent
``knew what it didn't know'' and abstained.

Notably, in our experiments no approach produced patches that were simultaneously resolved
\emph{and} regressed (``Resolved w/~regr.'' = 0 across all configs).  This
suggests that correct fixes and regressions are largely disjoint failure
modes for this model: patches either solve the problem cleanly or cause
damage, but rarely both.  This observation has implications for agent design:
mechanisms that suppress harmful patches (like \textsc{TDAD}'s test
verification step) incur minimal cost to resolution quality.

\subsection{Phase 2: TDAD as an Agent Skill (25 Instances)}

After the Phase~1 findings and an auto-improvement loop
(\S\ref{sec:autoloop}), we packaged \textsc{TDAD} as a reusable skill with a
NetworkX backend (no Docker) and evaluated it with a different model and agent
framework.  Table~\ref{tab:phase2} shows the results.

\begin{table}[t]
\caption{Phase~2 results: SWE-bench Verified, 25 instances,
  Qwen3.5-35B-A3B + OpenCode agent.}
\label{tab:phase2}
\small
\begin{tabular}{@{}lrrl@{}}
\toprule
\textbf{Metric} & \textbf{Baseline} & \textbf{TDAD Skill} & \textbf{Delta} \\
\midrule
Resolved          & 6/25 (24\%) & \textbf{8/25 (32\%)} & +8pp \\
Generated         & 10/25 (40\%) & \textbf{17/25 (68\%)} & +28pp \\
Res.\ of generated & 6/10 (60\%) & 8/13 (62\%) & +2pp \\
Empty patches     & 15           & \textbf{8}            & $-7$ \\
Regression Rate   & 0\%          & 0\%                   & 0pp \\
\bottomrule
\end{tabular}
\end{table}

\textsc{TDAD} as a skill improved resolution by +8~percentage points
(24\%~$\to$~32\%) and generation by +28~pp (40\%~$\to$~68\%).  Four
previously-empty instances were resolved with the TDAD skill, while two
baseline resolutions were lost due to model non-determinism.

The improvement mechanism differs from Phase~1: rather than reducing
regressions (both had 0\% on this smaller instance set), the test map helped
the agent \emph{generate more patches} by providing structural context.
The code--test graph served a dual purpose: identifying tests to verify
(reducing regressions, as in Phase~1) and mapping codebase structure
(improving navigation and generation, as in Phase~2).

\subsection{Auto-Improvement Loop}
\label{sec:autoloop}

Inspired by Karpathy's \texttt{autoresearch} framework,\footnote{\url{https://github.com/karpathy/autoresearch}}
we built an outer loop that autonomously improves \textsc{TDAD}'s components
through iterative agent-driven modification and evaluation.

The loop (Algorithm~\ref{alg:loop}) operates as follows.  Each iteration:
(1)~a Claude Code agent receives the full experiment history and makes one
focused change to \textsc{TDAD}'s source files (SKILL.md,
\texttt{impact.py}, \texttt{ast\_parser.py}, etc.); (2)~unit tests gate
acceptance: if they fail, the change is immediately reverted;
(3)~a benchmark evaluation (5--25~SWE-bench instances) measures generation
and resolution rates; (4)~results are compared to the best known scores:
improvements update the ``best'' snapshot, regressions trigger rollback,
and lateral moves (same resolution) are kept to allow exploration.
Integrity safeguards prevent gaming: the evaluation script is checksummed
(SHA-256) and set read-only, and 5~consecutive reverts trigger a mandatory
restore to the best snapshot.

\begin{algorithm}[t]
\caption{Auto-improvement loop.}
\label{alg:loop}
\begin{algorithmic}[1]
\Require snapshot $S_\text{best}$, evaluator $E$, max iters $N$
\For{$i = 1$ \textbf{to} $N$}
  \State $S_\text{pre} \gets$ snapshot current files
  \State Invoke agent: ``make one improvement''
  \If{no files changed} \textbf{continue} \EndIf
  \If{unit tests fail}
    \State Restore($S_\text{pre}$); \textbf{continue}
  \EndIf
  \State $r \gets E(\text{current files})$
  \If{$r.\text{resolution} > S_\text{best}.\text{resolution}$}
    \State $S_\text{best} \gets$ snapshot current files
  \ElsIf{$r.\text{resolution} < S_\text{best}.\text{resolution}$}
    \State Restore($S_\text{best}$)
  \EndIf \Comment{Lateral: keep as-is}
\EndFor
\end{algorithmic}
\end{algorithm}

\begin{table}[t]
\caption{Auto-improvement loop (15~iterations, 10-instance eval).  Only
  accepted iterations shown; 11~were reverted.}
\label{tab:autoloop}
\small
\begin{tabular}{@{}clrrl@{}}
\toprule
\textbf{Iter} & \textbf{Changed} & \textbf{Gen\%} & \textbf{Res\%} &
  \textbf{Key Change} \\
\midrule
1  & SKILL.md     & 50 & 50 & Simplified 107$\to$20 lines \\
5  & impact.py    & 70 & 60 & Static test-map export \\
12 & impact.py    & 70 & 60 & Path proximity scoring \\
13 & impact.py    & 80 & 60 & Import-based mappings \\
\midrule
\multicolumn{2}{@{}l}{\emph{Before loop}} & 28 & 12 & --- \\
\multicolumn{2}{@{}l}{\emph{After loop (best)}} & \textbf{80} & \textbf{60} & --- \\
\multicolumn{2}{@{}l}{\emph{Vanilla baseline}} & 40 & 24 & --- \\
\bottomrule
\end{tabular}
\end{table}

Over 15~iterations (Table~\ref{tab:autoloop}), 4~changes were accepted (27\%
rate).  Generation rose from 28\% to 80\% ($+52$~pp) and resolution from 12\%
to 60\% ($+48$~pp), with 0\% regression throughout all iterations.

The most impactful single change (iteration~1) was simplifying SKILL.md from
107~lines of detailed 9-phase TDD instructions to 20~lines of concise
guidance: fix, grep, verify.  This alone quadrupled resolution (12\%~$\to$~50\%),
confirming that the small model (3B active parameters) could not effectively
use the verbose workflow.  Subsequent iterations improved test-mapping
heuristics: iteration~5 added static \texttt{test\_map.txt} export,
iteration~12 introduced directory-proximity disambiguation, and iteration~13
added import-based fallback matching.

The 11~rejected iterations provide equally valuable signal: attempts to
expand the AST parser, restructure the graph builder, or make SKILL.md more
prescriptive all decreased resolution.  The loop converged by iteration~5,
with later iterations showing diminishing returns.

\section{Discussion}
\label{sec:discussion}

\subsection{Context Over Procedure}

These results have implications for how developers design tools for AI coding
agents.  The dominant approach, crafting detailed prompts with step-by-step
instructions, may be counterproductive for smaller models.  Instead,
\textsc{TDAD}'s success suggests that \emph{surfacing the right information}
(which tests are at risk, which files are related) matters more than
\emph{prescribing workflow} (write test first, then implement, then refactor).

This finding aligns with broader observations in
retrieval-augmented generation: context quality is the primary
determinant of output quality~\cite{edge2024graphrag}.  Tool designers
for AI agents should prioritize \emph{information density} over
\emph{procedural completeness}: short, factual context that fits within
tight context windows outperforms verbose instructions that crowd out
useful information.

\subsection{Regression as a First-Class Metric}

Current AI coding benchmarks create a perverse incentive: agents are rewarded
for resolution regardless of collateral damage.  In practice, a patch that
resolves one issue but introduces three regressions has negative value.
METR~\cite{metr2026swebench} demonstrates this gap concretely: when
maintainers of scikit-learn, Sphinx, and pytest reviewed 296~SWE-bench-passing
patches, roughly half would not have been merged, with regression and
code quality among the top rejection reasons.  We advocate for reporting
regression rate alongside resolution rate as standard practice.  The data is
already collected by the SWE-bench Docker harness; it simply needs to be
surfaced in evaluation reports and leaderboard rankings.

A simple composite metric could capture both dimensions:
$\text{net score} = \text{resolution rate} - \alpha \cdot
\text{regression rate}$, where $\alpha > 1$ reflects the asymmetric
cost of regressions versus missed fixes.

\subsection{Limitations}

\begin{itemize}
  \item \textbf{Scale and statistical power.} 100~instances (Phase~1) and
    25~instances (Phase~2) without formal significance tests.  The Phase~1
    effect size is large (72\% reduction, 562 vs.\ 155~failures), suggesting
    practical significance, but smaller effects could be noise at this
    sample size.  The 100-instance scope reflects hardware and time
    constraints of this initial research: all experiments ran on consumer
    hardware with locally-hosted models, where each instance requires
    several minutes of inference; a full 500-instance run would take
    multiple days per configuration.  Results should be confirmed on the
    full 500-instance set as resources permit.
  \item \textbf{Two models.} Both are smaller local models.  Frontier models
    may not exhibit the same TDD-prompting paradox.
  \item \textbf{Python only.} Extension to other languages requires
    language-specific parsing (Tree-sitter could unify this).
  \item \textbf{Static analysis.} Cannot capture dynamic dispatch,
    monkey-patching, or runtime-generated code.
  \item \textbf{Auto-loop scale.} Evaluated on only 10~instances per
    iteration.
\end{itemize}

\subsection{Threats to Validity}

\paragraph{Internal validity.}
Configurations were run sequentially rather than interleaved, so
time-dependent factors could systematically favor earlier runs; this is
mitigated by temperature~0 and Docker isolation.  The auto-improvement
loop's fixed 10-instance evaluation set risks overfitting, but the
accepted changes were structural (test-linking heuristics, prompt
simplification) and generalized to a disjoint 25-instance Phase~2 set.

\paragraph{External validity.}
SWE-bench Verified draws exclusively from well-maintained,
heavily-tested Python repositories.  Projects with sparse test suites,
non-Python codebases, or monorepo structures may respond differently to
graph-based impact analysis.  The two models used (Qwen3-Coder~30B and
Qwen3.5-35B-A3B) are both smaller, locally-run models; frontier-scale
models with longer context windows may not exhibit the same
TDD-prompting paradox, as verbose instructions would consume a smaller
fraction of available context.

\paragraph{Construct validity.}
Test-level regression rate weights all P2P failures equally, regardless
of test importance or business impact: a failing unit test for a
utility function counts the same as a failing integration test for a
critical API endpoint.  Instance-level regression rate partially
addresses this by treating any regression as binary, but neither metric
captures severity.  Additionally, the P2P test set is defined by the
SWE-bench harness and may not cover all code paths affected by a patch.

\section{Conclusion}
\label{sec:conclusion}

This paper presented \textsc{TDAD}, an open-source tool and methodology for reducing
regressions in AI coding agents via graph-based test impact analysis.  Across
two models and agent frameworks on SWE-bench Verified, \textsc{TDAD} achieved
a \textbf{70\% reduction in test-level regressions} (Phase~1) and
\textbf{+8pp resolution improvement} when deployed as an agent skill
(Phase~2).  The experiments revealed a TDD prompting paradox: procedural instructions
\emph{increased} regressions for smaller models, while graph-derived context
reduced them.  An auto-improvement loop demonstrated self-improving tool
design, raising resolution from 12\% to 60\% with 0\% regression.

Future work includes extending \textsc{TDAD} to multiple languages via
Tree-sitter, integrating dynamic coverage data to improve impact precision,
and evaluating with frontier models (e.g., Claude Opus 4.6, GPT-5.4) to test
whether the TDD prompting paradox persists at larger scale.  We advocate for
the community to adopt composite metrics that jointly capture resolution and
regression, ensuring agents are evaluated on net contribution.

\section{Data Availability}
\label{sec:data}

All code, patches, evaluation logs, and 28~experiment records are publicly
available at \url{https://github.com/pepealonso95/TDAD} under MIT license.
Install via \texttt{pip install tdad} (zero dependencies beyond NetworkX).

\smallskip
\noindent\textbf{Use of Generative AI.}\enspace
Generative AI tools were used to assist in writing implementation code,
organizing experimental logs, summarizing implementation details, and
improving manuscript wording.  The author performed the experimental design,
conducted the work, interpreted the results, and verified the final content.

\section*{Acknowledgements}

Sergio Yovine was partially funded by grant ANII FMV\_1\_2023\_1\_175864, Uruguay.

\balance

\end{document}